# Quantitative Assessments of Runway Excursion Precursors using Mode S data

Xavier Olive
ONERA / DTIS, Université de Toulouse
Toulouse, France
xavier.olive@onera.fr

Pierre Bieber
ONERA / DTIS, Université de Toulouse
Toulouse, France
pierre.bieber@onera.fr

*Abstract*— **A way to assess rare aircraft incidents (e.g., runway excursion) is to identify contributing factors (e.g., late braking, long landing, inappropriate flare, unstable approach) and to build a dependency tree (e.g., long landing may be the result of an unstable approach not followed by a go around) that describes the causality between these factors. Probabilities are then fed into such models in order to evaluate the assessed risk. When estimating such probabilities, many sources can be of interest. Airlines have access to the comprehensive flight data records of their fleet; manufacturers push to collect data for the aircraft they build; air traffic control log radar tracks. Albeit not as complete as other flight data records, Mode S data is very attractive, esp. for academics, as the data is open, may be published without obfuscation and offers reproducible results to the community. Mode S also provides an indiscriminate source of information (not limited to an airline or to an aircraft type) that is of great help for putting in context flights matching unusual patterns. We propose to discuss the advantages and limitations of an analysis based only on Mode S data with a case study around the runway excursion risk assessment.**

*Keywords— Air traffic safety, quantitative risk assessment, Mode S, ADS-B*

## I. Introduction

The Future Sky Safety (FSS) P4 project [1] aims at developing a prototype Risk Observatory that will assist in the safety assessment of the total aviation transport system. The Risk Observatory should help to compute the probabilities of occurrences of various incident and accident categories using risk models. As a contribution to this effort, Bieber et al. [2] introduced a risk model that identifies the contributing factors leading to runway excursion.

The frequencies of contributing factors shall be computed using data collected by stakeholders of the aviation transport system such as aircraft manufacturers, airline operators or Air Navigation Service Providers (ANSP). Limits were reached because quantitative evaluation of incidents and accidents would have required more information than was available from stakeholders. As necessary collected data are not shared openly, it has been difficult to compute the frequencies of contributing factors and, consequently, the probabilities of incidents and accidents. Also, the definition of contributing factors by experts is sometimes too vague for a proper numerical analysis.

As a way to tackle this limitation, we decided to investigate Mode S data, an alternative source of data that could be used in order to evaluate the frequencies of contributing factors leading to runway excursion. Mode S is a protocol for identification, localisation and communication between aircraft and surveillance equipment; it is mandatory in Europe and its full implementation is expected by 2020.

The OpenSky Network [4] provides an open access to their records for academic research, although its coverage is still behind commercial solutions like FlightRadar24. Mode S data is rather basic compared to the data collected by stakeholders of the aviation transport system, but they reveal another kind of precious information about the context where aircraft evolve that is not recorded elsewhere.

We explain in the following section the runway excursion model developed in [2], then describe the information present in Mode S data with the subset [5] that we use for our study. Further, we introduce a machine learning approach which we applied to a specific contributing factor for runway excursion. We show how the data-based statistical model we construct can 1) help to better specify contributing factors; 2) provide figures for quantifying the contributing factor we specifically addressed; and 3) challenge choices made when building the model. Finally, we mention limitations we encountered in the currently deployed implementation of this protocol as we tried to address other contributing factors of our model.

## II. Related Work and Literature Review

Aviation system stakeholders must collect relevant safety data in order to comply with regulations [6]. However, collected data cannot be easily shared as they are subject to privacy, warranty and competitiveness issues. Flight data monitors (FDM) collect data that can be a very rich source of information in terms of the number of features they provide; all analyses of the circumstances of accidents or incidents rely on such a comprehensive source of information.

In 2004, the European Aviation Safety Agency (EASA) and the Federal Aviation Administration (FAA) defined the Flight Operations Quality Assurance (FOQA) as a method of capturing, analysing and/or visualizing the data generated by an aircraft. Data sharing initiatives have also been sponsored by organisations in charge of the supervision of the aviation transport system. The ASIAS (Aviation Safety Information Analysis) programme [7] has established rules so that airline operators can share their data and preserve flight crew anonymity. The ASIAS programme would not authorize access to collected data that could easily be linked with a specific flight. Instead it provides to the ASIAS members general syntheses on the various categories of incident and accidents. In our case, general synthesis provided by ASIAS might be too general and not useful for computing the probabilities of contributing factors.

Previous works have already been conducted for safety analysis based on flight data. Das et al. [8] published results on anomaly detection based on NASA records, known as the Distributed National FOQA Archive (DNFA). This archive covers two million flights over 10 major carriers. They contain many continuous and discrete data from various on-board systems (propulsion systems, landing gears, cockpit switch positions, etc.), yet they do not offer a comprehensive view of the context in which aircraft evolve. Wang and Sherry [9, 10] also presented risk assessment analyses based on surveillance track data provided by the FAA National Offload Program. This data covers North America, and contains positional information at a sampling interval between 4 and 5 seconds.

In addition to safety assessments based exclusively on collected data, several groups developed risk models to deal with situations where data is not fully available. The Integrated Risk Picture model was developed by EUROCONTROL to assess accidents and incidents related with ANSP contributors [11]. This model was first used to compute baseline probabilities for accidents relying on collected data and predict these probabilities of accidents in the future. Here, the model no longer computes probabilities based on collected data; instead, it uses values to describe an assumed evolution of probabilities over time, taking into account evolution factors such as the growth of traffic or the planned introduction of new safety technologies in aircraft and ATM systems.

More recently, the Integrated Safety Assessment Model (ISAM) was developed following the same principles in order to analyse the NextGen framework of modernization of the US National Airspace [12]. Another risk model of interest is CATS (Causal Model for Air Transport Safety) developed by NLR [13]. In this model, the probabilities of some risk contributors could be provided by human analysis of accident and incident reports.

To our knowledge, Mode S data, thoroughly presented in Section IV has not yet been used for safety analyses. Although it exposes less information than what can be found in FDM or DNFA, it offers a continuous feed of data for analysis. It also provides information about context in which aircraft fly, that is usually not available in the data owned by flight operators. This source of information has recently been used for various applications, including modelling aircraft performance [14], and improving weather models [15, 16] and forecasts [17].

### III. MODELING THE RISK OF RUNWAY EXCURSION

A *backbone model* for the risk of runway excursion [2] was developed by partners of the Future Sky Safety (FSS) P4. This model manages in a consistent way contributing factors leading to a runway excursion. The top level view of the backbone model for the risk of longitudinal runway excursion is described in Fig. 1. It is a fault tree that relates failed situations such as "unstable approach" or "incorrect touchdown" and barrier failures such as "*4-Failure to manage stabilization in final approach*" or "*5.1- Absence of rejected landing*".

The fault tree notation [18] is also used to describe formally the relation between *Contributing factors* and barrier failures. A contributing factor can be a technical factor (e.g. airborne or ground equipment failure) or a human factor (flight crew, ATC, ground operator errors). Fig. 2 shows a part of the fault tree containing contributing factors leading to "*4. Failure to manage stabilization in final approach*". Contributing factors belong to the aircraft domain or to the ATM domain. Tab. I gives a list of aircraft domain contributing factors in category *Unstabilized Approach*. The fault tree also contains a combination of two contributing factors: "4.8 Crew request late change" that belongs to the aircraft domain and "3.8 ATC does not reject change request" that belongs to the ATM domain.

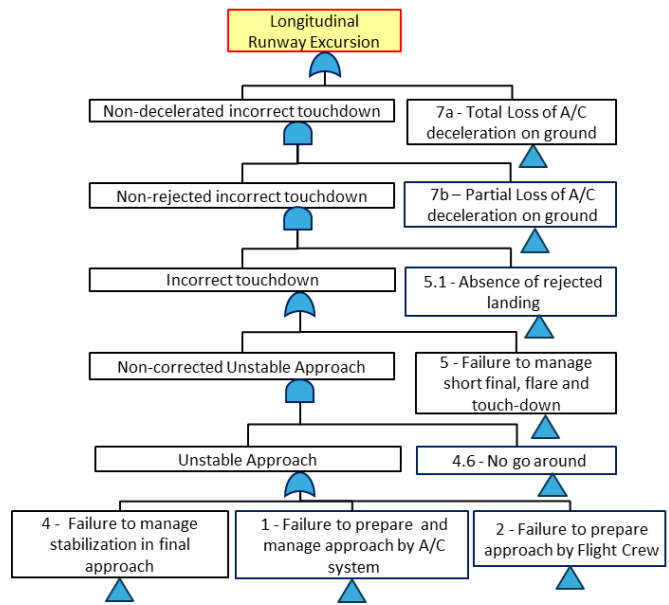

Figure 1. Top level view of the backbone model describing the risk of runway excursion.

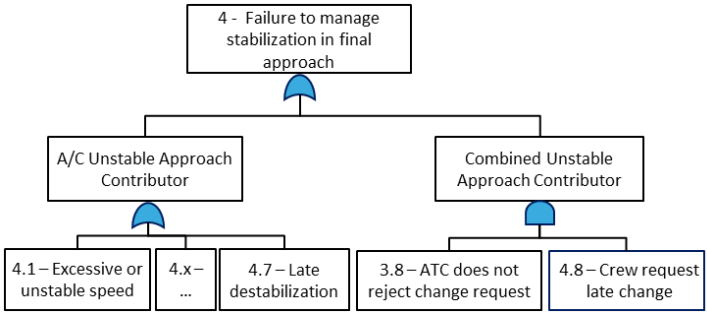

Figure 2. The fault tree for contributing factor *"4. Failure to manage stabilization in final approach"*

TABLE I. CONTRIBUTING FACTORS FOR UNSTABILIZED APPROACH

| Id | Contributing Factors | Probability |
|---|---|---|
| 4.1 | Excessive or unstable speed | $10^{-2}$ |
| 4.2 | Excessive or unstable lateral and vertical path | $3.10^{-2}$ |
| 4.3 | Excessive or unstable thrust | $10^{-2}$ |
| 4.4 | Late or inappropriate flaps/gear configuration | $5.10^{-3}$ |
| 4.5 | Inappropriate use of automation during approach | $10^{-3}$ |
| 4.6 | Absence of go around when unstable approach | $10^{-1}$ |
| 4.7 | Late destabilization of the approach | $10^{-2}$ |
| 4.8 | Crew requests late change | - |

The backbone model can be used to compute the probability of a failed situation given the probability of occurrence of each contributing factor [13]. The probability of occurrence of a contributing factor can be assessed using different sources.

For technical failures, we rely on failure rates computed by the aircraft manufacturer during the airworthiness certification process. For other contributors involving human errors, the aircraft manufacturer is not required by the regulation to compute probabilities. Average probabilities of contributing factors were established by an aircraft manufacturer using FDM collected data from several operators. These probabilities (occurrence rate per flight) are presented in Table I. However, this approach has several limitations.

Firstly, FDM data cannot directly be used to compute the probabilities for all potential contributors. For instance, FDM collected data did not provide the probability of contributor *"4.8 – Crew requests late change"* and the probability of other contributors in the ATM domain. Secondly, the probability of contributors is often based on hidden assumptions. For instance, to measure the probability of contributor *"4.1 - Excessive or unstable speed"*, we need to establish threshold values defining what is an excessive value for the aircraft speed during approach. A third limitation is that average probabilities do not provide useful information about the risk in a given context (e.g. the probability of runway excursion under given weather conditions at a given airport).

We investigated the use of Mode S data as a way to compute the probabilities of contributing factors that are used in the runway excursion backbone model. In particular, we wanted to assess whether it would be possible to overcome some of the three limitations that were presented previously.

IV. THE MODE S DATASET

Mode S has become one of the most important technologies in air traffic management as it supports the operation of secondary surveillance radar (SSR), traffic alert and collision avoidance systems (TCAS), and Automatic Dependent Surveillance–Broadcast (ADS-B). In practice, transponders in aircraft are selectively interrogated by sensors (radars) to provide situational awareness through the exchange of binary encoded information.

To be able to selectively interrogate aircraft, transponders aboard aircraft have been assigned a unique 24-bit identifier. The assignment of addresses is done by the national authority where the aircraft is registered. These identifiers, often referred as ICAO addresses, are included in all Mode S messages and identify each aircraft. We use in Section VI an aircraft database [25] to relate each ICAO address to an aircraft model.

Aircraft reply to ground sensor requests with messages of different types, called downlink formats (DF). We focus here on extended surveillance data from avionics such as intent- and status information via Comm-B messages (DF 20, 21). DFs 17 and 18 are different messages which are not transmitted upon interrogation. They contain all information needed to determine the aircraft's identity, location, and velocity. These *squittered* information are called Automatic Dependent Surveillance–Broadcast (ADS-B).

The OpenSky Network [4] is a crowd-sourced sensor network collecting air traffic data. Since ground-based receivers usually do not have a line of sight connection to the interrogators and since interrogations are done on a different frequency (1030 MHz), OpenSky only collects messages that are sent over the Mode S downlink on the 1090 MHz channel. The collected data used for this study contains ADS-B (DF 17, 18) and extended surveillance data (DF 20, 21) from February to July 2017. We trimmed our dataset to data collected by receivers based in Toulouse area.

Apart from callsigns and position coordinates, messages contain information about flight dynamics parameters. ADS-B provides *ground speed* (GS) information, i.e. the horizontal speed of the aircraft relative to the ground (usually displayed on passengers' entertainment systems). GS is the vector sum of the *true air speed* (TAS) and the current wind, to be decomposed along a headwind and a crosswind components. Pitot tubes measure the difference between static and dynamic pressures in order to compute an *indicated air speed* (IAS). IAS is based on the sea level standard atmospheric density, hence differs from the TAS, that is the relative velocity between the aircraft and the surrounding air mass.

When an aircraft points its nose in a direction, the angle relative to North is the *heading angle*; its actual path travelled on the ground, because of the wind, makes a different angle known as *track angle*. When an aircraft lands, it must align its track angle with the *magnetic bearing* of the runway. Depending on the direction of landing, the two parallel runways at Toulouse–Blagnac airport have a magnetic bearing of 324° (QFU 32) and 144° (QFU 14).

In this paper, we focus on several types of messages (Tab. II). We get positional (latitude, longitude, GPS and barometric altitude) and velocity information (track angle, ground speed and vertical rate) from ADS-B data (DF 17) received at a rate of about one message per second. We also focus on some specific Comm-B messages (DF 20, 21), only sent upon request, namely BDS 5,0 messages (*track and turn report*), which include roll and track angle; and BDS 6,0 messages (*heading and speed report*), with heading and indicated air speed information.

TABLE II.  DATA PARAMETERS ACCORDING TO DOWNLINK FORMAT

| DF 17 (ADS–B) | DF 20, 21 (Comm-B) | |
| --- | --- | --- |
| callsign | callsign | BDS 2,0 |
| latitude (°) | roll angle (°) | BDS 5,0 |
| longitude (°) | track angle (°) | BDS 5,0 |
| GPS altitude (ft) | ground speed (kts) | BDS 5,0 |
| barometric altitude (ft) | true air speed (kts) | BDS 5,0 |
| vertical rate (ft/min) | heading (°) | BDS 6,0 |
| ground speed (kts) | indicated air speed (kts) | BDS 6,0 |
| track angle (°) | mach number | BDS 6,0 |

## V. ASSESSMENT OF CONTRIBUTING FACTORS

This section addresses the conjunction of contributing factors *"4.8 Crew requests late change"* and *"3.8 ATC does not reject change request"*. Current regulations do not allow any access to recorded communications between ATC and pilots. However, an analysis of trajectory patterns may give strong hints of how pilots and ATC interacted. Although this approach does not enable us to detect a rejected late change from the data, we show in this section how we may detect specific late change requests which got clearance from the ATC. For the sake of clarity, we will write in this section *"Crew requests late change"* in place of the conjunction of contributing factors *"4.8 Crew requests late change"* and *"3.8 ATC does not reject change request"*.

### A. Mathematical context

Trajectories are mathematical objects used to describe the evolution of a moving object. They are described by a state vector with parameters $(x(t), y(t), \cdots)$ that evolve in time. In practice, this state vector is only known at some sampled times. For clarity concerns, we will name trajectory a sequence of recordings associated to an aircraft. The explosion of recorded data makes the study of trajectories a popular topic and opens new fields of research in data mining common patterns [3, 19, 20] and identifying outliers in a set of trajectories.

Unsupervised learning is about inferring the hidden structure from *unlabelled* data. Instead of having labels in part of our data, e.g. information about hazardous situations, we must focus on a set of unlabelled trajectories, try to grasp their structure and find outlying elements. Principal Component Analysis (PCA) [21, 22] is a tool for exploratory data analysis which explains the variance in the data. It cuts down the complexity of data by projecting data samples on axes holding the more variance, determined by an eigenvalue decomposition of the covariance matrix of our samples.

Functional Principal Component Analysis (FPCA) [23] is a useful tool to analyse functional data, such as altitude, speed, or track angle profiles present in the data. Considering a dataset of n functions $x_1(t), x_2(t), \cdots x_n(t)$, (e.g. representing an altitude profile), we generalise the decomposition over eigenvectors holding the more variance, as defined by Principal Component Analysis (PCA), to the infinite dimension. FPCA results from the Karhunen-Loève decomposition of a signal $x(t)$:

$$x(t) = \sum_{i=1}^{+\infty} \theta_i \gamma_i(t) \qquad (1)$$

where the principal component scores $\theta_i = \langle \gamma_i, x \rangle$ are centered and uncorrelated random variables s.t. $\mathbb{E}(\theta_i) = \lambda_i \geq 0$. We can interpret the random scores $\theta_i = \langle \gamma_i, x \rangle$ as proportionality factors that represent strengths of the projection of each individual trajectory on the i-th principal component function.

FPCA provides eigenfunction estimates that can be interpreted as modes of variation. They offer a visual tool to assess the main directions in which functional data vary. Nicol [24] then explains how FPCA can be equivalent to a PCA over discretisations of our $x_i(t)$ functions as a series of variables defined at each given time.

### B. Analysis of the track angle profiles

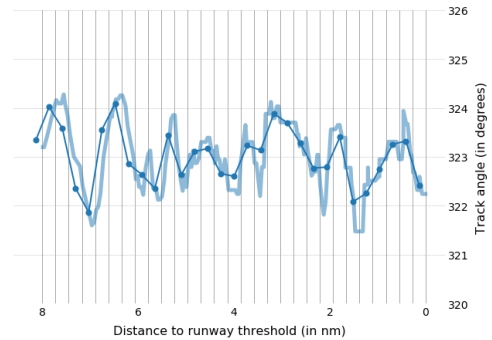

Figure 3. The track angle profile of Flight AF100GN on February 21 is resampled on the last 8 nautical miles before the runway threshold.

We focus on a set of 2194 trajectories landing at Toulouse airport in February 2017. Among them, 1398 trajectories landed on 32R and 32L runways: we trimmed their final approach trajectory to the last 8 nautical miles before the runway threshold. We arbitrarily chose a resampling in 30 equidistant points on the 8 nautical miles before the runway threshold (Fig. 3). Eq. (1) becomes for each $x_i(t)$ signal as a linear combination over a set of 30 eigenfunctions ($\gamma_1, \gamma_2, \cdots$):

$$x_i(t) = \sum_{j=1}^{30} \theta_{i,j} \gamma_j(t) \quad (2)$$

The resulting 30 component vectors are then passed through a PCA engine, resulting in a variance ratio distribution shown on Fig. 4. The diagram reads as follows: the first component holds 35 % of the total variance, the second component about 17 %, etc.; and all variance ratios sum to 1. In our case, the total variance is distributed on the few first components; we can choose to ignore components with a variance ratio under a given threshold and reduce our functional data to a handful of $\theta_{i,j}$ components.

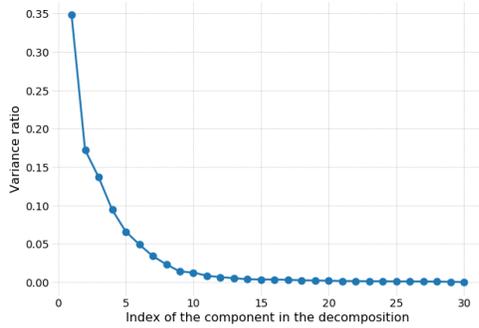

Figure 4. Explained variance ratios for each component of a FPCA decomposition over the 1398 track angle profiles.

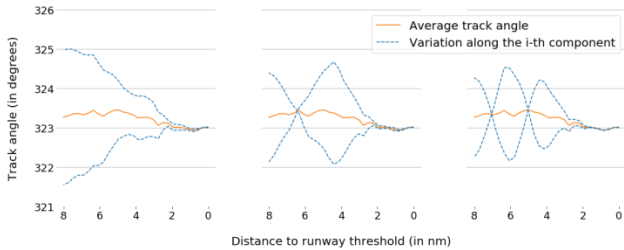

Figure 5. FPCA decomposition of the track angle profile and its decomposition over the three first eigenfunctions with the more variance.

Fig. 5 shows the three first eigenfunctions $\gamma_1(t)$, $\gamma_2(t)$ and $\gamma_3(t)$ holding more than 10 % of the total variance over our set of 1398 track angle profiles. They describe the three most significant modes of variation present in our data. The full line represents the average signal and the dashed lines variations of $\pm 3\sqrt{\lambda_j}$ in the *direction* of $\gamma_j(t)$. This means that we can now rewrite each track angle profile $x_i(t)$ as a decomposition:

$$x_i(t) = \theta_{i,1}\gamma_1(t) + \theta_{i,2}\gamma_2(t) + \theta_{i,3}\gamma_3(t) + \cdots \quad (3)$$

that is $\theta_{i,1}$ along the first mode of variation, $\theta_{i,2}$ along the second mode of variation, etc.

We can interpret the modes of variation as follow: the first component $\theta_{i,1}$ quantifies a variation around the average profile (full line) and between the two dashed lines. This mode depicts a variation on the distance to the roll-out point, i.e. the point where the aircraft aligns with the runway.

$\theta_{i,2}$ and $\theta_{i,3}$ are particularly interesting as they depict low frequency variations on the track angle profile during final approach. It may seem surprising to see such variations in the track angle profiles at this stage, when the aircraft should align its ground trajectory with the magnetic bearing of the runway.

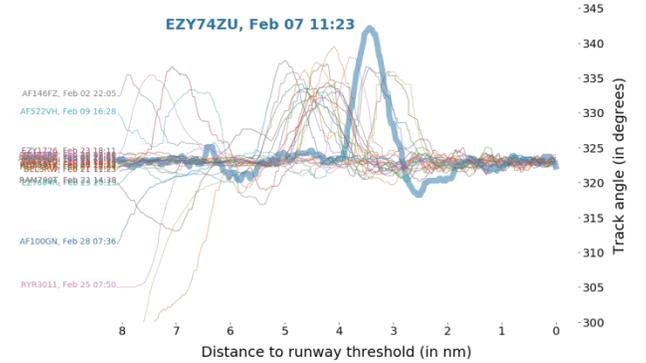

Figure 6. Heading signals for aircraft landing on QFU 32 with $|\theta_{i,2}| > q_2^{0.98}$

Let $q_j^{0.98}$ be the 98th percentile of the $\{|\theta_{i,j}|\}$ projections of $x_i(t)$ on $\gamma_j$ and select on Fig. 6 all track angle profiles with the most significant components $\theta_{i,2}$ along $\gamma_2(t)$:

$$|\theta_{i,2}| > q_2^{0.98} \quad (4)$$

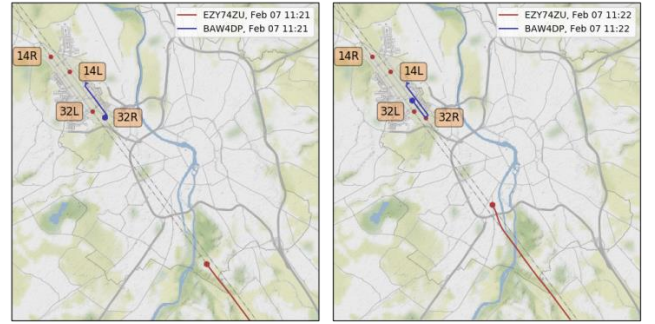

Figure 7. Between 11:21 and 11:22 on February 7, EZY74ZU shifts from its alignment on 32L to 32R, after BAW4DP takes off.

Fig. 7 shows context and suggests an interpretation to what Eq. 4 reflects: EZY74ZU on February 7 was aligned on runway 32L at 11:21; at that time, BAW4DP was ready for take-off. One minute later, BAW4DP started rolling to take-off, then EZY74ZU switched and aligned to runway 32R. Toulouse airport terminal is actually located on the east side of the runway, and landing on 32R saves a significant time of taxiing, which matters for commercial flights. EZY74ZU most probably asked for a late runway change during approach.

At this point, we should work with experts in order to select a subset of trajectories (e.g. one month); define a proper threshold in place of $q_2^{0.98}$ and $q_3^{0.98}$ so as to be able to properly extract all flights requesting a late runway change in final approach; compute the probability for a larger set of trajectories (e.g. one year, fixed, or on a sliding time window); and eventually, inject the probability in the fault-tree and compute a runway excursion risk.

*C. Cross-analysis with altitude profiles*

A similar FPCA analysis can be performed on barometric altitude profiles: a similar interpretation suggests that the second and third components reflect low frequency variations around the glide path (GP), to be related to contributing factor *"4.2 Excessive or unstable lateral or vertical path"*.

TABLE III. NUMBER OF TRAJECTORIES WITH EQ. 4 VERIFIED.

|  | $q_2^{0.98}$ | $q_3^{0.98}$ | $q_2^{0.98}$ or $q_3^{0.98}$ |
|---|---|---|---|
| heading | 27 | 27 | 51 (3.6 %) |
| altitude | 27 | 27 | 51 (3.6 %) |
| heading and altitude |  |  | 6 |

Tab. III counts the number of flights requesting a runway change, based on the projections on $\gamma_2(t)$ and $\gamma_3(t)$, and those showing signs of excessive or unstable vertical path. 51 track angle profiles (out of 1398, i.e. 3.6 %) reflect a late runway change; among them, 6 trajectories (out of 51, i.e. 11.8 %) also show signs of an unstable approach. Fig. 8 shows how one of the 6 flights extracted from 0, asks for a late runway change and does not follow a stable glide path.

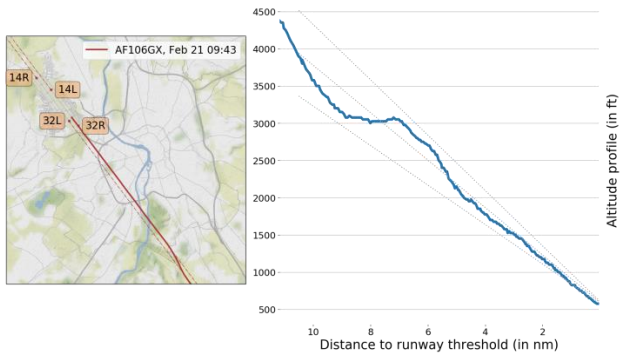

Figure 8. On February 21, AF106GX shifts from 32L to 32R and presents an unstable glide path (altitude in feet). The dashed lines on the altitude profile depict the $3 \pm 0.5°$ glide path.

A model for evaluating these contributing factors, i.e. *"Crew requests late change"* or *"Excessive or unstable vertical path"*, consists of the FPCA transformation matrix (how to project each track angle profile on the $\gamma_i(t)$ functional coordinates), and of threshold values to be validated with experts. As a reference, we computed the probability on six months of data with $q_2^{0.98}$ and $q_3^{0.98}$ as threshold values: among 387 trajectories asking for a late runway change, 64 trajectories also show signs of excessive or unstable vertical path, with an estimated probability of 16.54 %.

As a matter of fact, this possible correlation may lead us to reconsider the classical independence assumption used to compute probabilities of failed situations in the fault tree of Fig. 2. Further investigations may lead us to reconsider the fault tree and introduce a new contributor accounting for the combination of contributing factors *"Crew requests late approach"* and *"Excessive or unstable vertical path"*.

VI. LIMITS OF MODE S DATA FOR SAFETY ASSESSMENTS

*A. Speed reports*

A common rule of thumb for qualifying non stabilised approaches is to consider aircraft flying too high and too fast on their glide path. Previous section studies how to work with track angle and altitude profiles. Considering speed signals could also bring insight to our analysis.

DF 17 (ADS-B) provides ground speed signals. Fig. 9 plots the distribution of ground speeds according to the aircraft model in the last minute of their airborne trajectories. The distributions show the importance of considering aircraft types when studying speed profiles. Even between aircraft of similar types and sizes (Airbus 319, 320 and 321), the distribution is centred on different ground speeds. Indeed, the piloting during approach depends on the aircraft model as each type of aircraft is limited by its stalling speed.

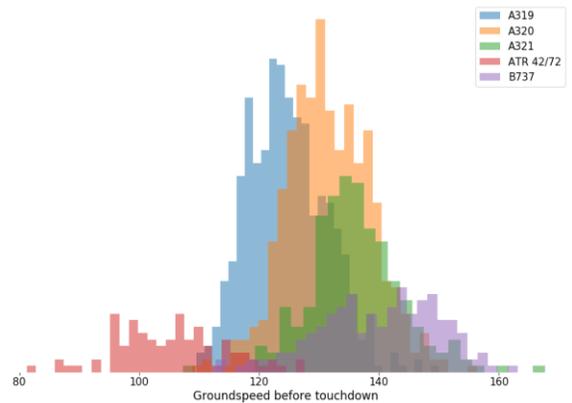

Figure 9. Average ground speed distribution on the last nautical mile before landing according to the aircraft type (limited to most frequent aircraft)

It would be difficult to apply the presented approach on speed profiles: ground speed is mainly used by the FMS to estimate its time of landing but not used for piloting; indicated air speed is more relevant (it appears in the cockpit) as it is related to the aerodynamics of the aircraft.

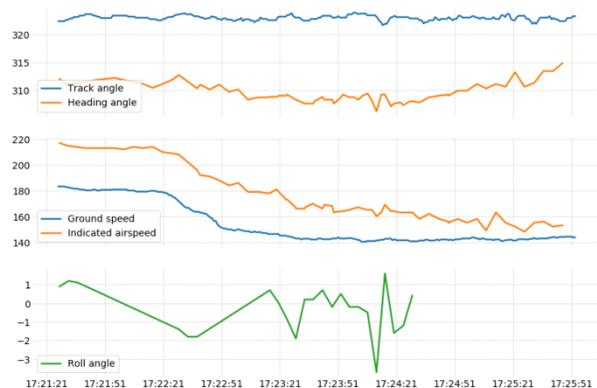

Figure 10. Flight EZY158T attempts to land at Toulouse airport on February 5. The plots suggest strong gusty crosswind.

DF 20, 21 (EHS), namely BDS 6,0 messages (heading and speed report), provide profiles of heading and indicated airspeed which call for further analysis. Fig. 10 shows the example of Flight EZY158T on February 5, a day of strong gusty crosswind (see the influencing factor related to weather conditions). The gap between the track and heading angles is a sign of a crab approach; the roll angle profile shows a lateral instability during approach; and the stable profile of GS to be compared with the more hectic profile of IAS shows how IAS is also impacted by the wind gusts during final approach.

Future works should include a model of wind per altitude layer to be maintained from BDS 4,4 messages (meteorological routine air-report): these messages are already being used in building meteorological model [15]. Then, we could focus on the difference between IAS and wind, an indicator to what the pilot tries to maintain, i.e. a stable airspeed regardless of the wind gusts, and closely relates to the total energy of the plane.

Unlike ADS-B, EHS messages are only sent upon request from the ATC. Mode S being expected to be fully implemented in Europe by 2020, we found that it is still difficult today to rely on EHS data as for many flights landing in Toulouse, ATC requests for such messages look erratic. Fig. 11 shows the average number of BDS 6,0 messages per aircraft during final approach according to the day and time in February 2017: in addition to a probable maintenance of the ATC system on February 10 between 11am and 8pm, our study is hampered by the lack of messages received in the second half of the month.

Discrepancies in the number of requests for each type of EHS messages are to be expected between control centres, though BDS 6,0 message rate should be of one per second. Apart from the observed maintenance issue, several factors may explain the lack of messages we observe: apart from CRC checks on received messages, decoding DF 20, 21 messages requires assumptions on the types of messages that have been requested. As things stand, we do not receive nor decode request messages, which forces us to discard messages when data looks consistent with many kinds of BDS messages.

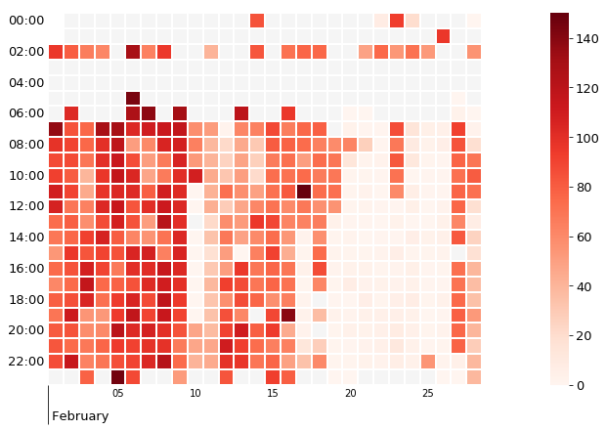

Figure 11. Average number per flight of BDS 6,0 messages during final approach. This calendar view very different profiles from day to day.

### B. GPS precision and touchdown point

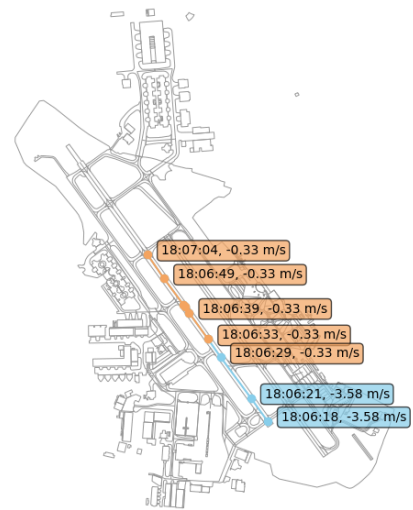

Figure 12. ADS-B data prove to be not precise enough for properly determining the touchdown point on the runway. (Flight DLH83K on Feb. 25)

Another topic of interest when analysing the risk of runway excursion would be to determine the touchdown point on the runway. Equipped aircraft have a sensor which detects the pressure on the landing gear when the aircraft lands, which triggers a different type of DF 17 positional messages. Fig. 12 plots an aircraft trajectory, with *surface* messages as brown dots and *airborne* messages as blue dots. The label next to it reads the vertical speed, also available in DF 17: the colour associated to the label changes when vertical speed stabilises and suggests that the aircraft has landed. Looking at both sources of data, we can only suppose that the aircraft touched the runway between 18:06:21 and 18:06:33, that is a 12-second confidence interval, not enough for determining a precise touchdown point.

## VII. Conclusion

This paper presented a model for the risk of runway excursion and a ML approach to assess specific contributors to this risk from real traffic Mode S data. The model was constructed using one month of data around Toulouse airport, then was used on a longer period to compute figures that could be fed into our model or could challenge hypotheses of non-correlation that were made when building the model.

The proposed approach addresses three limitations identified in Section III: 1) we showed how to compute the probabilities of contributors that are not covered by other sources of data (FDM, communications with ATC). But our investigation also showed that using Mode-S data has currently some drawbacks such as missing data due to an imperfect infrastructure for receiving and decoding messages; 2) the approach helps to uncover hidden assumptions that may have been made by experts. In particular, the unsupervised learning approach attempts to quantify what experts would label as "excessive"; 3) context helps refining the analysis that could have been done with FDM data, in addition to offering the possibility to compare flight profiles landing at a given airfield under the same conditions, even if onboard data is not owned by the same stakeholder.

It is our belief that using Mode S data together with other sources of data is a reasonable approach to quantify safety requirements for air traffic. Its full implementation in Europe expected by 2020 should make it an unavoidable source of data for safety analysis.

## VIII. Acknowledgements

This study has received funding from the *European Union's Horizon 2020 Research and Innovation Programme* under Grant Agreement no. 640597.